\documentclass[11pt]{article}
\usepackage{jheppub}
\usepackage{graphicx}
\usepackage{amsmath,amssymb,amstext,amsfonts} 
\usepackage{mathrsfs}
\usepackage[hang]{subfigure}
\usepackage[utf8]{inputenc}
\usepackage[a4paper,top=1.89in, bottom=0in, left=2in, right=0in]{geometry}
\usepackage{xcolor}
\usepackage{braket}
\usepackage{array}
\usepackage[bottom]{footmisc}
\usepackage{tikz}
\usepackage{tabularx}
\usepackage{setspace}
\usepackage{enumitem}
\usepackage[numbers]{natbib}

\usepackage{color}

\newcommand{\be}{\begin{equation}}
\newcommand{\ee}{\end{equation}}
\newcommand{\ba}{\begin{eqnarray}}
\newcommand{\ea}{\end{eqnarray}}

\numberwithin{equation}{section}

\begin{document}
\title{Phase Transitions in 4D Gauss-Bonnet-de Sitter Black Holes}

\author[1]{Gareth Arturo Marks}
\author[1,2]{Fil Simovic}
\author[1,2]{Robert B. Mann}

\affiliation[1]{Department of Physics and Astronomy, University of Waterloo,\\
	Waterloo, Ontario N2L 3G1, Canada}
\affiliation[2]{Perimeter Institute for Theoretical Physics,\\
	31 Caroline St. N.,Waterloo, Ontario N2L 2Y5, Canada}
\emailAdd{fil.simovic@gmail.com}
\emailAdd{gamarks@uwaterloo.ca}
\emailAdd{rbmann@uwaterloo.ca}

\abstract{We investigate thermodynamic aspects of black holes in the recently formulated four dimensional Gauss-Bonnet theory of gravity, focusing on its asymptotically de Sitter ($\Lambda>0$) solutions. We take a Euclidean path integral approach, where thermodynamic data is fixed at a finite radius `cavity' outside the black hole to achieve equilibrium in the presence of the cosmological horizon. Working in the extended phase space where the cosmological constant is treated as a thermodynamic pressure, we study the phase structure of both uncharged and charged solutions, uncovering a wealth of phenomena. In the uncharged case, black holes are found to undergo either the standard Hawking-Page-like transition to empty de Sitter space or a small-large transition (akin to those seen in charged AdS black holes in pure Einstein gravity) depending on the pressure. We also find a triple point where the radiation, small, and large black hole phases coexist, and a zeroth-order phase transition within a narrow range of Gauss-Bonnet coupling parameter. Reentrant phase transitions between radiation and a small black hole also exist inside a certain parameter range. Similar phenomena are found in the charged case, with the charge parameter playing a role analogous to the Gauss-Bonnet coupling parameter in determining the phase structure.}


\maketitle

\section{Introduction}

The subject of black hole thermodynamics has a long history, dating back to observations in the 1970's by Bardeen, Carter, and Hawking that variations in the mass of a black hole obey a law closely resembling the first law of thermodynamics for ordinary systems \cite{bardeen1973a}. Since then, advances in our understanding of general relativity, symmetries in gauge theory, and quantum field theory have led to concrete interpretations of the variables that enter into the first law of black hole mechanics. With a natural temperature and entropy being associated with a black hole in the semi-classical setting \cite{bekenstein1973a,hawking1975a}, much work has since been done to uncover the thermodynamic properties of black hole spacetimes. This is motivated in part by the belief that the thermodynamic description arises from a coarse-graining of some microscopic degrees of freedom (which are poorly understood in the case of gravity) and by the fact that black holes display a wide variety of thermodynamic phenomena that are interesting in their own right. Historically, special attention has been paid to asymptotically anti-de Sitter (AdS) black holes, which possess a natural thermodynamic equilibrium via the attractive effective potential of AdS and have an important dual interpretation in AdS/CFT in terms of thermal CFT states \cite{maldacena1999b}. One of the most notable features of such black holes is their ability to undergo phase transitions resembling those of ordinary systems, the most familiar example being the Hawking-Page transition between an empty spacetime and a large black hole. This transition in the bulk is dual to a thermalization transition in the boundary CFT, and has played an important role in our understanding of the so-called black hole information paradox \cite{hawking1983a,harlow2016}. 
\\

Over the years, studies concerning the phase structure of AdS black holes have revealed a rich thermodynamic landscape, with reentrant phase transitions, triple points, and superfluid-like transitions being observed for black holes in a variety of contexts \cite{kubiznak2012,altamirano2013,frassino2014,hennigar2017f}. This phase structure is not only interesting in its own right, but allows us to probe the phase structure of the strongly coupled dual CFTs and reveal new phases that are otherwise invisible to perturbation theory (e.g. triple points in the QCD phase diagram \cite{cui2021}). The extent to which these phenomena persist in the asymptotically {\it de Sitter} (dS) case is being slowly revealed, as new techniques are developed for studying black hole thermodynamics in spacetimes with positive $\Lambda$. Such black holes are the natural ones to study, both in their potential applications to finite-boundary holography (like dS/CFT), and because recent measurements indicate black holes in our universe are at least approximately asymptotically de Sitter \cite{perlmutter1999,percival2010,planckcollaboration2019}. 
\\

Yet, the inclusion of positive $\Lambda$ introduces difficulties not present in the AdS case. The system is manifestly out of equilibrium due to the presence of the cosmological horizon, and defining the asymptotic mass is not straightforward like it is in asymptotically flat and AdS spaces. To overcome these issues, we take a Euclidean path integral approach, where thermodynamic data is fixed on a finite radius surface in the spacetime. This not only makes equilibrium manifest for the system, but also allows for the straightforward determination of thermodynamic quantities associated with the spacetime through the relationship between the partition function and Euclidean action. Such an approach has been used recently to understand a wide variety of asymptotically de Sitter spacetimes, with interesting new phenomena appearing \cite{simovic2019a}.
\\

In this paper, we will investigate a new class of black hole solutions that appear in the recently formulated\footnote{A number of inequivalent formulations have emerged, only one of which is the focus of this paper. See Section \ref{theory} for further details.} 4D Gauss-Bonnet gravity \cite{glavan2020,hennigar2020a,fernandes2020}. Gauss-Bonnet gravity is an extension of general relativity that supplements the Einstein-Hilbert action with a curvature-squared term of the form
\be\label{Lgb}
\mathcal{L}_{GB}=R^{2}-4 R^{\mu \nu} R_{\mu \nu}+R^{\mu \nu \rho \sigma} R_{\mu \nu \rho \sigma}\ .
\ee
This theory has enjoyed significant theoretical interest owing to the fact that such a Lagrangian arises naturally as the leading correction to GR in 10D gauged supergravity and as a quantum correction in heterotic string theory \cite{cecotti1988}. It is also the leading correction appearing in the Lovelock class of higher-derivative gravity theories, which have been explored extensively over the years. Ordinarily, the Gauss-Bonnet term is purely topological in $D=4$ rendering the full theory indistinguishable from GR. Recently however, there has been considerable effort in reformulating Gauss-Bonnet gravity in four dimensions, 
(analogous to what was done for Einstein gravity in two dimensions  \cite{mann1993})
such that non-trivial dynamics emerge \cite{glavan2020,lu2020,hennigar2020a,fernandes2020}. Such a theory would represent the only competitor to GR as a metric theory of gravity in 4D whose equations of motion are second-order (being free of ghost instabilities). A number of formulations have emerged, which over a short period of time have enjoyed considerable interest. Already many aspects of 4D Gauss-Bonnet gravity and its black hole solutions have been studied, including their shadows, tidal effects, causal structure, quasinormal modes, gravitational lensing, and more \cite{li2021,dadhich2020,konoplya2020,islam2020,clifton2020,panah2020}. Moreover, a variety of other solutions (with familiar analogues in Einstein gravity) have also been constructed, including cosmic strings, wormholes, and rotating black holes \cite{lin2020,liu2020,kumar2020}.
\\
\\

What remains unexplored however is the thermodynamic nature of asymptotically {\it de Sitter} black hole solutions in this new theory. While anti-de Sitter thermodynamics has been examined already \cite{wei2020,mansoori2021,wang2021,hegde2021}, this work represents the first attempt at uncovering thermodynamic features of the $\Lambda>0$ solutions of 4D Gauss-Bonnet gravity. Focusing in particular on the phase structure of these black holes, we will find a host of interesting phenomena including Hawking-Page, small-large, and reentrant phase transitions, as well as triple points and zeroth order phase transitions, many of which ordinarily only appear in higher dimensional theories of gravity or when auxiliary charges are present.
\\

This paper is organized as follows: In Section \ref{cavity}, we briefly describe the Euclidean path integral approach to studying the thermodynamic properties of black hole solutions in de Sitter space. In Section \ref{theory} we describe the origin and formulation of Gauss-Bonnet gravity in four dimensions, and evaluate the bulk and boundary actions of the theory. In Section \ref{uncharged}, we examine uncharged black hole solutions to the theory, and study their phase structure through the equilibrium temperature and free energy. In Section \ref{charged} we consider the charged solutions. We summarize our results and comment on future extensions of this work in Section \ref{conclusions}. Natural units where $\hbar = G = c = 1$ are used throughout.

\section{Euclidean Thermodynamics}\label{cavity}

We will use Euclidean methods to study the thermodynamic properties of various solutions of the 4D Gauss-Bonnet theory. Such methods are based on the quantum mechanical path integral, which expresses the transition amplitude of a quantum system in terms of an integration over all possible paths between the system's initial and final states $\ket{q_i}$ and $\ket{q_f}$, with each path being weighted by the exponential of the action for the quantum field $q$:
\be
\braket{q_i(t_i)|q_f(t_f)}=\int\mathcal{D}[q] e^{-iI/\hbar}\ .
\ee
In the context of the gravitational field, the path integral formally defines a quantum theory of gravity and has been used extensively in the study of gravitational thermodynamics \cite{hawking1978,hamber2009}. In this application, one exploits the fundamental relationship between the classical {\it Euclidean} action $I_E$ of the theory and the quantum mechanical partition function $\mathcal{Z}$,
\be
F=-T\log \mathcal{Z}\approx TI_E
\ee
where $F$ is the free energy of the system and $T$ is the equilibrium temperature. The above relation arises from a semi-classical and saddle point approximation of the full path integral, and encodes naturally the Kubo-Martin-Schwinger (KMS) condition for thermal Green's functions through the periodicity $\beta$ in the imaginary time coordinate $\tau=i t$ of the Euclidean section \cite{fulling1987}. With access to the partition function, one can straightforwardly determine the equilibrium thermodynamic properties of the system through the usual formulas from statistical mechanics
\be\label{thermoquants}
E=\frac{\partial I_E}{\partial \beta}\ ,\qquad S=\beta\frac{\partial I_E}{\partial \beta}-I_E\ ,
\ee
where the quantities $E$ and $S$ have the interpretation as being the mean thermal energy and entropy of quantum fields (including the gravitational field) in the given spacetime. 
\\

This Euclidean approach offers many advantages to the usual methods, especially in its application to asymptotically de Sitter spacetimes. In de Sitter space, the presence of the cosmological horizon (which generally radiates at a temperature different from that of any black hole) introduces a heat flux that places the black hole system out of equilibrium. By fixing boundary value data in the path integral at a finite-radius boundary within the spacetime, equilibrium is manifest in the system and the issues caused by the presence of the cosmological horizon are resolved. This has the physical interpretation of isolating the black hole within an isothermal cavity, where heat flux at the surface is allowed such that the temperature is fixed to the redshifted Hawking temperature of the black hole. Specifically we have that $T=\beta_c^{-1}$, where $\beta_c$ is the locally observed periodicity in imaginary time $\tau$ at the location of the cavity $r=r_c$. This $T$ coincides with the ordinary Hawking temperature of the black hole, blueshifted to the location of the cavity. Such an approach was first applied to the study of asymptotically flat black holes \cite{braden1990}, and has since been used extensively to understand the thermodynamic properties of a variety of asymptotically de Sitter systems \cite{carlip2007,simovic2019,haroon2020,simovic2021}.  
\\

As is common in the discussion of the phase structure of black hole systems, we treat the cosmological constant as a thermodynamic pressure according to the relation
\begin{equation}\label{pressure}
P=-\frac{\Lambda}{8\pi}\ ,
\end{equation}
where $P<0$ when $\Lambda$ is positive. In a Hamiltonian analysis, the variation of $\Lambda$ manifests as a supplementary pressure-volume term entering the ordinary first law of black hole thermodynamics
\be\label{ext}
dM=TdS+VdP+\Omega\,dJ\ ,
\ee
where the left-hand side now has the interpretation of an enthalpy variation $d H$ instead of an internal energy variation $d U$ \cite{kastor2009b}. In this extended phase space, many new and interesting thermodynamic features appear, including Van der Waals-like transitions, superfluid and reentrant phase transitions, and more \cite{kubiznak2016}. It has become somewhat standard to study the phase structure of black holes in this extended phase space, owing to two main facts. The first is the natural appearance of a $\Lambda$-variation in many cosmological and string theory contexts, for example in the Brown-Teitelboim mechanism, in cosmological models  incorporating a decay law for $\Lambda$, and in anisotropic cosmologies \cite{bousso2008}. The second is that many phenomena that  occur even at fixed $\Lambda$ are revealed only when its variation is incorporated into the thermodynamic description, making the extended phase space represented by \eqref{ext} the natural one in which to operate.

\section{Gauss-Bonnet Gravity in Four Dimensions}\label{theory}

The usual formulation of Gauss-Bonnet gravity arises from the $n=2$ truncation of the full Lovelock Lagrangian
\be\label{lovelock1}
\mathcal{L}=\frac{1}{16 \pi }\sqrt{-g} \sum_{n=0}^{n_{\max}} \alpha_{n} \mathcal{L}^{(n)}\ ,
\ee 
where $\alpha_n$ is the coupling of the $n$-th term and $\mathcal{L}^{(n)}$ is the 2n-dimensional Euler density, given by a series of contractions involving higher powers of the Riemann tensor:
\be\label{lovelock2}
\mathcal{L}^{(n)}=\frac{1}{2^{n}}\,\delta_{c_{1} d_{1} \ldots c_{n} d_{n}}^{a_{1} b_{1} \ldots a_{n} b_{n}}\, {R_{a_{1} b_{1}}}^{c_{1} d_{1}} \dots {R_{a_{n} b_{n}}}^{c_{n} d_{n}}
\ee
Lovelock theory has played an important role in understanding the implications of modifications to general relativity, being the most general metric theory of gravity that is (i) symmetric, (ii) ghost-free, and (iii) manifestly diffeomorphism-invariant \cite{lovelock1971}. In Lovelock theory, $\mathcal{L}^{(0)}=1$ with $a_0=-2\Lambda$, and $\mathcal{L}^{(1)}=R$, so that the Einstein-Hilbert action is recovered when the sum is truncated at $n=1$.  Importantly, $n_{\text{max}}=\text{Floor}[(D-1)/2]$, as the $\mathcal{L}^{(n)}$ term only contributes to the equations of motion when $D>2n$. As a result, Lovelock gravity is equivalent to general relativity when $D=4$, and the first non-trivial correction is obtained only when $D=5$, so that $\mathcal{L}^{(2)}$ enters in both the action and the field equations. The resulting theory is known as Gauss-Bonnet gravity, and the new term entering the Einstein-Hilbert action is given by \eqref{Lgb} above.
\\

Recently, significant attention has been given to the formulation of Gauss-Bonnet gravity in four spacetime dimensions. According to Lovelock's theorem \cite{lovelock1971}, Einstein gravity is the unique metric theory of gravity satisfying conditions (i)-(iii) when $D=4$. However, a number of recent studies have shown that it is possible to obtain a sensible $D\rightarrow 4$ limit of Gauss-Bonnet gravity in such a way that the curvature-squared term \eqref{Lgb} contributes non-trivially to the theory, while retaining the desirable properties (i)-(iii) dictated by Lovelock's theorem. These explorations were initiated by the work of \cite{glavan2020a}, who demonstrated the existence of a non-trivial four-dimensional limit to 5D Gauss-Bonnet gravity, at the level of the equations of motion. This initial formulation has since been shown to suffer from a number of issues \cite{gurses2020,ai2020,arrechea2020,arrechea2021}, and has been superseded by other constructions which yield a non-trivial 4D Gauss-Bonnet theory at the level of the action \cite{lu2020}. One way to accomplish this is to start with the higher-dimensional theory and perform a dimensional reduction along with an appropriate rescaling of the Gauss-Bonnet coupling so that the ordinarily singular part of the limit yields a finite, non-trivial contribution to the lower dimensional action. Another method involves a conformal rescaling trick \cite{hennigar2020a,fernandes2020}, based on one used in two-dimensional gravity \cite{mann1993},
having the advantage of not relying on a particular choice of higher-dimensional geometry. The resulting theory, and the one we focus on in this paper, is a scalar-tensor Horndeski-type theory given by the following bulk action:
\be\label{action0}
I=\int d^{4} x \sqrt{-g}\left[R-2 \Lambda+\alpha\left(\phi\, \mathcal{L}_{GB}+4 G^{a b} (\partial_{a} \phi) (\partial_{b} \phi)-4(\partial \phi)^{2} \square \phi+2\left((\partial \phi)^{2}\right)^{2}\right)\right]
\ee
Here, $\alpha$ is the Gauss-Bonnet coupling parameter, $G^{ab}$ is the Einstein tensor, and $\phi$ is an auxiliary scalar degree of freedom that can be seen to arise from either a compactification of the higher-dimensional Gauss-Bonnet theory, or from the conformal rescaling employed in \cite{hennigar2020a} to obtain the four-dimensional action \eqref{action0}.
\\

For our purposes we require the Euclideanized version of \eqref{action0} supplemented by appropriate boundary terms arising from the fixed boundary data at the cavity. This yields
\begin{small}
	\begin{align}\label{fullact}
	I_{\rm E} &=  -\int_{\cal M} d^4 x \sqrt{g}  \left\{\frac{1}{16 \pi } \left[R-2 \Lambda + \alpha\left( \phi\mathcal{L}_{GB}+ 4G^{\mu \nu}\partial_\mu \phi \partial_\nu \phi - 4(\partial \phi)^2 \Box\phi +2((\partial\phi)^2)^2 \right) \right]- \frac{1}{4}F_{\mu\nu}F^{\mu\nu} \right\}
	\nonumber\\
	&
	-\frac{1}{8 \pi } \int_{\partial \mathcal{M}}\!\! d^{3} x \sqrt{\gamma} \left[K  -\frac{2}{3}\alpha\phi(K^3 - 3KK^{(2)}+2K^{(3)})+2\alpha (K^{\mu\nu}\partial_\mu \phi \partial_\nu \phi +Kn^\mu n^\nu\partial_\mu \phi \partial_\nu \phi - K(\partial \phi)^2)  \right]  
	\nonumber\\
	&-\int_{\partial \mathcal{M}}d^3x\sqrt{\gamma}F^{\mu\nu}n_\mu A_\nu\ ,
	\end{align}
\end{small}
\noindent \hspace{-11pt} where the boundary terms were recently computed in \cite{ma2020}. 
This action corresponds to the compactification of $D$-dimensional Gauss-Bonnet gravity to four dimensions on a $(D - 4)$-dimensional maximally symmetric space of zero curvature, after taking the $D \rightarrow 4$ limit. $g_{\mu\nu}$ is the full spacetime metric, while $\gamma_{ij}$ is the induced metric on the boundary surface (defined by an outward-pointing unit normal vector $n^{\mu}$), with $g$ and $\gamma$ being their respective determinants (see  \cite{hennigar2020a} for more details). $K_{\mu \nu} = {\gamma_\mu}^\rho \nabla_\rho n_\nu$ the second fundamental form, and we further define
\be
K = K^\mu_{\;\; \mu}\ ,\qquad K^{(2)} = K^{\mu}_{\;\;\nu}K^{\nu}_{\;\;\mu}\ ,\qquad K^{(3)} = K^{\mu}_{\;\;\nu}K^{\nu}_{\;\;\lambda}K^{\lambda}_{\;\;\mu}\ .
\ee

Being interested in charged solutions, we have also included a $U(1)$ Maxwell field in the action with an electromagnetic boundary term appropriate for the fixed-charge ensemble we are considering here. The terms appearing in the first line of \eqref{fullact} are the bulk terms from which the equations of motion are derived. The boundary terms ensuring a well-posed Dirichlet problem appear in the second and third lines. In addition, we will be considering a constant-$r$ hypersurface in a (Euclidean) spherically symmetric geometry. In this case, the induced metric takes the form
\be 
\gamma_{ij} dx^i dx^j = f(r) dt_{\rm E}^2 + r^2 d\theta + r^2\sin^2\theta d\phi \, .
\ee
For the bulk metric and scalar field we have the following ansatz
\be\label{ansatz}
ds^2 = f(r) d\tau^2 + \frac{dr^2}{f(r)} + r^2 d\Omega^2_	2 \ , \qquad \phi = \phi(r)\ ,
\ee
where $d\Omega_2^2$ is the metric on $\mathcal{S}^2$. The solution for the metric function is known to have two branches, given by 
\be\label{metric}
f_{\pm}(r) = 1 + \frac{r^2}{2\alpha}\left(1 \pm \sqrt{1 + \alpha\left(\frac{4\Lambda}{3} + \frac{8 M}{r^3}\right) }\ \right)\ .
\ee
In this paper we consider the $f_{-}$ branch of the metric function, as it approaches the ordinary Schwarzchild de Sitter solution in the limit $\alpha \rightarrow 0$. In addition, the solution for the scalar field is either $\phi(r) = \log\left[(r - r_0)/l\right]$ for some integration constants $r_0$ and $l$ or 
\be\label{phi}
\phi_\pm(r) =  \int \frac{\sqrt{f(r)}\pm 1}{r\sqrt{f(r)}}dr\ .
\ee
We will consider the solution $\phi_-(r)$ only. This choice  falls off as $1/r$, suitable for  spacetimes of constant asymptotic curvature,
whereas the $\phi_+(r)$ branch diverges logarithmically at large $r$. We take the $U(1)$ gauge field $A_{\mu}$ to be spherically symmetric, with only the radial component being non-zero and taking the form
\be
A_r(r) = -\frac{1}{4}\left(\frac{Q}{r}-\frac{Q}{r_+}\right)\ .
\ee
We note that in the charged case, the boundary term associated with the Maxwell field exactly cancels an additional bulk contribution. Thus, the explicit charge dependence will drop out from our final expression for the action, and thermodynamic quantities will differ in the charged case only by the replacement of the metric function \eqref{metric} by its charged analogue.

\subsection{Evaluating the Bulk Action}

Deriving thermodynamic quantities for the theory under consideration amounts to the evaluation of the on-shell Euclidean action \eqref{fullact} and the application of the relations \eqref{thermoquants}. Rather than directly integrating \eqref{fullact}, we will proceed by using the field equations in the 4DGB theory to rewrite the bulk action as a total derivative. We begin by substituting the spherically symmetric ansatz \eqref{ansatz} into the bulk Lagrangian to obtain the following expression
\begin{align}\label{simp1}
r^2\mathcal{L}_{\mathrm{bulk}} &= \,2\,(1-f-\Lambda r^2  - 2rf')- r^2f'' + \alpha \left[2r^2 f^2(\phi ')^4 - \left(\frac{4}{3}r^2ff'+ \frac{16}{3}rf^2\right)(\phi')^3  \right. \nonumber\\ 
&\left.+ \,4f\left(rf' + f - 1\right)(\phi')^2- 4f'(f - 1)\phi ' + \frac{\partial}{\partial r}\left( -\frac{4}{3}r^2f^2(\phi ')^3 + 4f'(f - 1)\phi   \right) \right]
\end{align}
where primes indicate differentiation with respect to $r$ and it is understood that $f=f(r)$. Note that that the overall factor of $r^2$ is what the square root of the determinant $\sqrt{g}$ will contribute to the radial integral in the bulk action. Therefore, it is this term that we want to express as a derivative so that we may evaluate the bulk action. First, we note that the Einstein part can easily be written as a total derivative:
\begin{align}
2\,(1-f-\Lambda r^2  - 2rf')- r^2f'' = \frac{\partial}{\partial r}\left(  2r - \frac{2}{3}\Lambda r^3 - 2fr - r^2f'\right)\ 
\end{align}
We must now re-express the remaining terms in the Gauss-Bonnet part of the action in a similar way. To do this, we substitute the solution \eqref{phi} into terms $\phi' f$ appearing in \eqref{simp1}. After some simplification we obtain
\begin{align}\!\!\!\! r^2\mathcal{L}_{\mathrm{bulk}} = \frac{\partial}{\partial r}\left(
2r - \frac{2}{3}\Lambda r^3 - 2rf - r^2f' + \frac{2\alpha}{r}\left[ 1 - 2\sqrt{f} - f^2 + 2f^{3/2} + 2rf'(f - 1) \phi\right]
\right)
\end{align}
The radial integration covers the entire Euclidean section, and is therefore taken from the event horizon $r_+$ to the cavity radius $r_c$. Before progressing further, we make a note regarding the scalar field $\phi(r)$. The solution \eqref{phi} is given in terms of an indefinite integral, which cannot be evaluated analytically. Only the Euclidean section of the spacetime contributes to the path integral, so we choose to integrate \eqref{phi} from $r_+$ to $r$ so that our solution for the scalar field is explicitly
\be\label{phi2}
\phi(r) =  \int_{r_+}^{r} \frac{\sqrt{f(r)}- 1}{\sqrt{f(r)}}dr\ .
\ee
This must be integrated numerically whenever $\phi(r)$ is not evaluated at $r=r_+$. There is a freedom in the choice of value for $\phi(r_+)$ however, since this simply amounts to a shift by a constant value in the $\alpha\phi\mathcal{L}_{GB}$ term in \eqref{fullact}. Such a shift only results in a constant offset in the total entropy $S$, leaving other thermodynamic quantities unaffected \cite{mir2019,ciambelli2020}. In light of this, we choose for convenience
\be\label{phicond}
\phi(r_+)=0\ ,
\ee
which further normalizes the entropy to simply take on the Bekenstein-Hawking value of $S=\pi r_+^2$. With this, we can express the bulk (Euclidean) action as 
\be
I_E^{\mathrm{bulk}} = -\frac{\beta}{4}\left(2r - \frac{2}{3}\Lambda r^3 - 2rf - r^2f' + \alpha\left[ \frac{2}{r} - \frac{4}{r}\sqrt{f} - \frac{2}{r}f^2 + \frac{4}{r}f^{3/2} + 4f'(f - 1) \phi\right]\right)\bigg\rvert_{r = r_+}^{r = r_c}\nonumber
\ee
Here, $\beta$ is the periodicity in imaginary time $\tau$ required to avoid a conical singularity in the $\tau-r$ plane at the bifurcation surface, and the locally observed periodicity at the location of the cavity is $\beta_c = \sqrt{f(r_c)}\,\beta $. Therefore, equilibrium is achieved by fixing the temperature at the cavity to be $T=\beta_c^{-1}$.

\subsection{Evaluating the Boundary Action}

We will now compute the boundary action. The Euclidean metric \eqref{ansatz} is a smooth manifold at the horizon, with topology $\mathbb{R}^2 \times \mathbb{S}^{2}$. We therefore require a boundary term only at the location of the cavity. To compute the boundary term we follow \cite{haroon2020}, using the orthonormal projectors 
\be
\tau_\mu^\nu = \delta_\mu^t \delta^\nu_t  \qquad  \rho_\mu^\nu = \delta_\mu^r \delta^\nu_r 
\qquad \sigma_\mu^\nu =  \delta_\mu^\theta \delta^\nu_\theta + \delta_\mu^\phi \delta^\nu_\phi
\ee
to decompose the curvature into temporal, radial, and angular parts (there is no sum over $r,$ $\tau$, $\theta$, or $\phi$). The orthogonal projectors satisfy the following relations: 
\be
\tau_\mu^\lambda \tau_\lambda^\nu = \tau_\mu^\nu \, , \quad \rho_\mu^\lambda \rho_\lambda^\nu = \rho_\mu^\nu \,,\quad \sigma_\mu^\lambda \sigma_\lambda^\mu = \sigma_\mu^\nu \, ,
\ee
and
\be
\tau_\mu^\nu \tau_\nu^\mu = \rho_\mu^\nu \rho_\nu^\mu = 1  \, , \quad \sigma_\mu^\nu \sigma_\nu^\mu = 2\ .
\ee
The boundary term is composed of various traces and contractions of the extrinsic curvature tensor $K_{ij}$. For a constant-$r$ surface, the extrinsic curvature computed for the outward-pointing unit normal vector $n^{\mu}$ is
\be
K_i^j =  \frac{f'}{2 \sqrt{f}} \tau_i^j + \frac{\sqrt{f}}{r} \sigma_i^j \, ,
\ee
where $k$ measures the curvature of the constant time slices of the boundary. Using these definitions we obtain the following:
\begin{align}
K  &=  \frac{1}{2 \sqrt{f}} \left(f' + \frac{4f}{r}  \right) \, ,
\nonumber\\
K^{(2)} &= \left(\frac{f'}{2 \sqrt{f}} \right)^2 + 2 \left(\frac{\sqrt{f}}{r} \right)^2 \, ,
\nonumber\\
K^{(3)} &= \left(\frac{f'}{2 \sqrt{f}} \right)^3 + 2 \left(\frac{\sqrt{f}}{r} \right)^3 \, .
\end{align}
Finally, we recall that the outward-pointing normal one-form is
\be 
n_\mu dx^\mu = \frac{dr}{\sqrt{f(r)}}
\ee
Using these results we can readily compute the boundary action, finding that
\be 
I_E^{\text{boundary}} = -\frac{\beta}{2}\left[2\,r_c f(r_c)+\frac{1}{2}r_c^2f'(r_c) - 2\alpha f(r_c)f'(r_c)\phi(r_c) \right]\ .
\ee

\section{Uncharged 4D Gauss-Bonnet Black Holes}\label{uncharged}

We will now study the thermodynamic properties of $4$-dimensional uncharged Gauss-Bonnet-de Sitter black holes. Combining the bulk and surface terms we have already computed, we obtain the following expression for the total action:

\begin{align}\label{act1}
I_E&=\frac{\beta}{4}\left[2(r_+ - r_c) +\frac{2}{3}\Lambda(r_c^3 - r_+^3) -2r_cf(r_c) - r_+^2f'(r_+) \right. 
\\ &\ \ \ +\left.\alpha\left(\frac{2}{r_+} - \frac{2}{r_c} + \frac{4\sqrt{f(r_c)}}{r_c} - \frac{4f(r_c)^{3/2}}{r_c} + \frac{2f(r_c)^2}{r_c} +4\Big(f'(r_c)\phi(r_c)-f'(r_+)\phi(r_+)\Big)\right) \right] \nonumber
\end{align}
It is now standard to normalize the action so that the empty de Sitter spacetime has zero energy and entropy. This is accomplished by subtracting from \eqref{act1} the same action evaluated for empty de Sitter space, with the bulk metric matched to that of $I_E$ at the boundary surface $r=r_c$ (this reference action will be denoted $I_S$). To do this, we simply evaluate \eqref{act1} with the replacement $f(r)\rightarrow f_0(r)$, where $f_0(r)$ is the metric function evaluated for $M = 0$, and rescale the $d\tau'$ variable appearing in the integration by enforcing that $g_{\text{\,Schw-dS}}$ and $g_{\text{\,dS}}$ are equal at $r=r_c$. The bulk radial integration in the background term extends now from 0 to $r_c$, covering the entire Euclidean section. The result is that
\begin{align}
I_S = &  \frac{\beta}{4}\left[-2r_c +\frac{2}{3}\Lambda r_c^3 -2r_cf_0(r_c) \right. 
\\ & \qquad + \left.\alpha\left(- \frac{2}{r_c} + \frac{4\sqrt{f_0(r_c)}}{r_c} -\frac{4f_0(r_c)^{3/2}}{r_c} +\frac{2f_0(r_c)^2}{r_c} +4f_0'(r_c)\phi_0(r_c)\right) \right]\ ,\nonumber
\end{align}
where $\phi_0$ is the same branch of the scalar field solution \eqref{phi2} evaluated using $f_0(r)$.
The full action we wish to consider is then
\begin{align}
I_* =I_E-I_{S}\ .
\end{align}
With this action we compute the energy and entropy in the standard way using \eqref{thermoquants}. This requires first making the substitutions $f'(r_+) = 4\pi\sqrt{f(r_c)}/\beta_c$ and $\beta = \beta_c / \sqrt{f(r_c)}$, as the variations in \eqref{thermoquants} are done with respect to the equilibrium (cavity) temperature. We find that the thermal energy and entropy are:
\begin{align}
E = \frac{\partial I_*}{\partial \beta_c}&= \frac{r_c}{2} -\frac{\Lambda}{6}r_c^3 +r_c\frac{f_0(r_c)}{2} - r_c\frac{f(r_c)}{2} + \frac{\Lambda(r_c^3 - r_+^3)}{6\sqrt{f(r_c)}}-\frac{(r_c - r_+)}{2\sqrt{f(r_c)}}
\\ \nonumber &\ \ \   +\alpha \left( \frac{3}{2r_c}  -\frac{f(r_c)}{r_c} +\frac{f(r_c)^{3/2}}{2r_c} + \frac{f'(r_c)\phi(r_c)}{\sqrt{f(r_c)}} + \frac{1}{2\sqrt{f(r_c)}}\left(\frac{1}{r_+} - \frac{1}{r_c}\right)
\right.  \\ \nonumber &\ \ \ 
\left.  - \frac{\sqrt{f_0(r_c)}}{r_c}  + \frac{f_0(r_c)^{3/2}}{r_c} - \frac{f_0(r_c)^2}{2r_c} - f_0'(r_c)\phi_0(r_c)
\right)\ ,
\nonumber\\
S = \beta_c E - I_* &= \pi r_+^2+4\pi \alpha\phi(r_+) 	\ .
\end{align}
Note that the entropy differs from our expectation in the Schwarzschild case only by a term proportional to  $\phi(r_+)$. This term vanishes in light of our choice of boundary value \eqref{phicond} for the scalar field; we leave it in simply to show explicitly the effect a different choice would produce. The entropy is then precisely that of the ordinary Schwarzschild black hole in Einstein gravity. We note that the phase structure of these black holes is largely unaffected by a different choice of boundary condition, with only numerical changes occurring, but with the same qualitative features. Finally, the temperature $T=\beta_c^{-1}$ is simply
\be
T = \frac{f'(r_+)}{4\pi\sqrt{f(r_c)}}
\ee
which is just the standard Hawking temperature of the black hole blue-shifted to the cavity radius.

\subsection{The First Law}

The (extended) first law of thermodynamics  reads
\be\label{firstlaw}
dE=TdS+VdP+\sigma dA+\Phi_{GB}\,d\alpha
\ee
for uncharged Gauss-Bonnet black holes.
As mentioned above, the pressure--volume term $VdP$ appears since we are considering variations in the cosmological constant in the extended phase space according to the relation \eqref{pressure}. $V$ is the thermodynamic volume of the system, which in general differs from the geometric volume of the black hole. Additionally, a work term $\sigma dA$ associated with changes in the cavity size $A$ is present, where $\sigma$ is interpreted as the surface tension of the cavity. Finally, the $\Phi_{GB}d\alpha$ term must be included to account for variations in the Gauss-Bonnet coupling.
\\

As the 4D Gauss-Bonnet theory is a scalar-tensor theory, it is natural to ask whether a further generalization of the first law is required due to the presence of the scalar field $\phi$. Formally, we may consider adding a term representing the variation of the scalar charge along with its conjugate potential, so that
\be\label{firstlaws}
dE=TdS+VdP+\sigma dA+\Phi\, d\phi\ ,\qquad \Phi\equiv\left(\dfrac{\partial E}{\partial \phi}\right)\ .
\ee
In the context of string theory \cite{gibbons1996a}, such a first law appears generically for black holes as both their mass and area depend on the values of the moduli fields $\phi_{\infty}$ at spatial infinity. These moduli label different ground states of the theory, as the string coupling $g_s$ is related to the vacuum expectation value of the dilaton $\phi$ at infinity. The moduli do not correspond to new conserved charges however, as they are not associated with a new integration constant \cite{hajian2017}. The question of whether or not an inclusion of the form \eqref{firstlaws} is warranted is thus subtle. This issue was recently explored in \cite{astefanesei2018a}, where black holes in Einstein-Maxwell-dilaton gravity were considered, and it was shown that neither the Smarr formula nor the first law required the addition of a scalar charge term. For our purposes, we can therefore take the extended first law to read
\be
dE=TdS+VdP+\sigma dA+\Phi_{GB}\,d\alpha\ ,
\ee
Having determined the energy $E$, temperature $T$, and entropy $S$, we can determine the conjugate variables $\{V,\lambda,\Phi_{GB}\}$ using the first law \eqref{firstlaw}. The explicit forms of the conjugate variables are available in Appendix A. In \figurename{ \ref{fig:Fig1}} we display two of them, the thermodynamic volume $V$ and the cavity surface tension $\lambda$, with the shaded areas representing regions in the parameter space where the corresponding quantity is positive.

\begin{figure}[h]
\centering\includegraphics[width=0.5\textwidth]{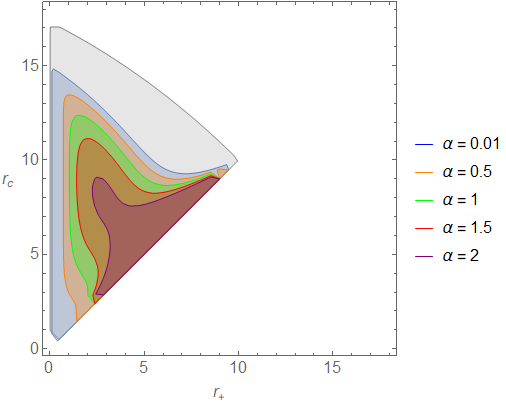}\includegraphics[width=0.5\textwidth]{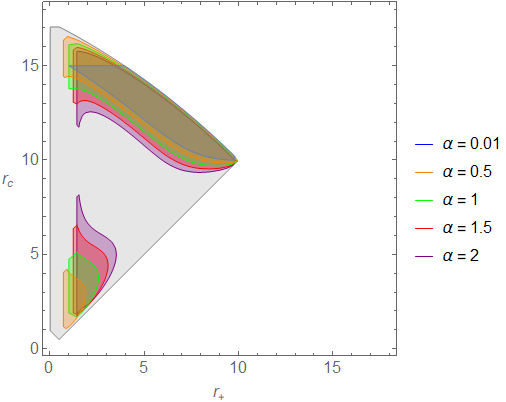}
	\caption{ Values of $r_c$ and $r_+$ for which the thermodynamic black hole volume and surface tension are positive with $P = -0.00040$. \textbf{Left}: The outermost grey region is that in which $V$ is real, and is independent of $\alpha$. The inner colored regions show the points for which $V > 0$ for various $\alpha$ as indicated by the legend. \textbf{Right:} The grey region shows where the surface tension $\sigma$ is real (which coincides precisely with the grey region on the left) while the colored regions show where it is positive.}
	\label{fig:Fig1}
\end{figure}

In general, the thermodynamic volume $V$ differs from the geometric `volume' of the black hole as defined by its areal radius $r_+$. As can be seen in the left figure above, for certain choices of Gauss-Bonnet parameter $\alpha$ the thermodynamic volume is actually negative. One may worry that in such areas $V$ cannot have the interpretation of a physical volume, however, in such regions the minus sign can simply be accounted for by a change in sign of the work term $VdP$. In these (unshaded) areas then, the thermodynamic volume can retain its positivity with the interpretation that work is now done by the cavity on the system, rather than work being done on the cavity. This is important to keep in mind owing to the fact that such `negative' thermodynamic volumes appear rather generically in both asymptotically dS and AdS black holes \cite{simovic2019,simovic2019a,xu2020}.
\\

On the right of \figurename{ \ref{fig:Fig1}}, we show in a similar way regions where the cavity surface tension $\lambda$ is positive. Outside of these regions we have $\lambda<0$, so that $\lambda$ in fact becomes a surface pressure rather than a tension. In both cases, the quantities are positive for any size black hole when the Einstein limit $\alpha\rightarrow 0$ is taken, as indicated by the grey shaded wedge. The diagonal boundary marks the point where $r_+=r_c$, representing the maximum possible black hole size, while the upper curved boundary represents the point where the cosmological horizon intersects the cavity.

\subsection{Phase Structure of Uncharged Black Holes}

\begin{figure}[h]
	\includegraphics[width=0.49\textwidth]{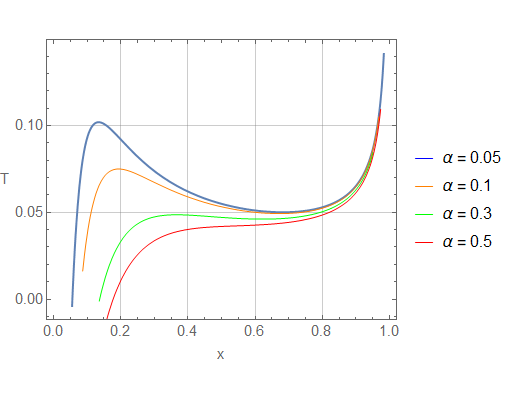}\quad\includegraphics[width=0.52\textwidth]{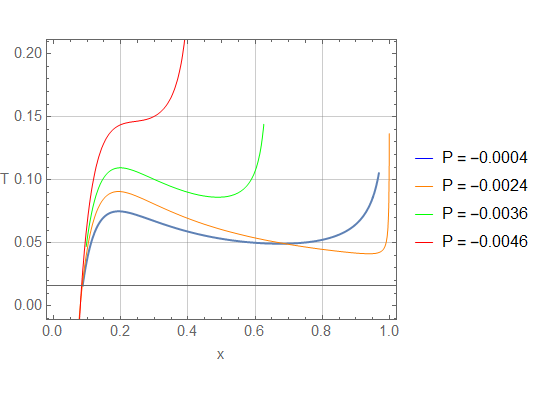}
	\caption{Temperature $T$ as a function of $x=r_+/r_c$ for fixed cavity radius $r_c=4$  \textbf{Left:} Varying Gauss-Bonnet coupling with $P = -0.004$. 
		\textbf{Right:} Varying pressure with $\alpha= 0.1$.}
	\label{fig:Fig2}	
\end{figure}

Having determined the energy, entropy, and temperature of the 4D Gauss-Bonnet black hole solutions, we can begin to study their phase structure. The key quantity is the free energy $F = E  - TS$,  which is globally minimized by the equilibrium state of the system. The order parameter here is the horizon radius $r_+$, which uniquely distinguishes black holes of different sizes as well as the empty de Sitter phase (corresponding to $r_+=0$). To begin with we note that in general, phase transitions within the system occur at locations where multiple distinct horizon radii possess the same temperature. At such points there will be multiple black hole phases with differing sizes (including possibly the empty de Sitter phase) and with differing free energies. In such cases, the thermodynamically preferred state will be the one with the lowest free energy. There may also be inflection points in the temperature where it remains single-valued, signalling the merging of a large and small black hole at a second-order phase transition. Whether these multiple competing phases represent valid solutions to the field equations cannot however be determined by the temperature function alone, requiring additional checks to ensure such phases are indeed accessible by the system. With this in mind, we begin by plotting in \figurename{ \ref{fig:Fig2}} the temperature $T$ as a function of the (rescaled) event horizon radius $x=r_+/r_c$ for fixed Gauss-Bonnet coupling $\alpha$ and various pressures $P$ (i.e. various $\Lambda$), as well as fixed $P$ and various $\alpha$ for the uncharged black hole solution given by \eqref{metric}. 
\\

First note that the transition of $T(r_+)$ from the single- to multi-valued regime occurs both at fixed $P$ when the Gauss-Bonnet coupling $\alpha$ is large enough, and at fixed $\alpha$ when the pressure is large enough. These critical values are shown in the red curves of \figurename{ \ref{fig:Fig2}} for a particular choice of cavity radius. For small $\alpha$ and $P$, $r_+$ is multivalued as a function of the temperature, indicating possible phase transitions. Thus, we expect that there is a compact region of $(\alpha,P)$-space where $r_+$ is a multivalued function of $T$, hence where first-order phase
transitions are possible. In contrast, outside of these regions, there is generally only one thermodynamically allowed state. Finally, note that as the pressure becomes increasingly negative ($\Lambda$ becomes large) a limiting size appears in the figure on the right, where no black hole solutions can be constructed with $r_+\sim r_c$. In these cases, the limit on the size of the black hole corresponds to the point where the cosmological horizon reaches the cavity radius. As we must always have $r_{\text{cosmo}}>r_c$, $T(r_+)$ terminates at this endpoint, before the event horizon reaches the cavity wall. 
\\

\begin{figure}[h]
	\;\;\includegraphics[width=0.49\textwidth]{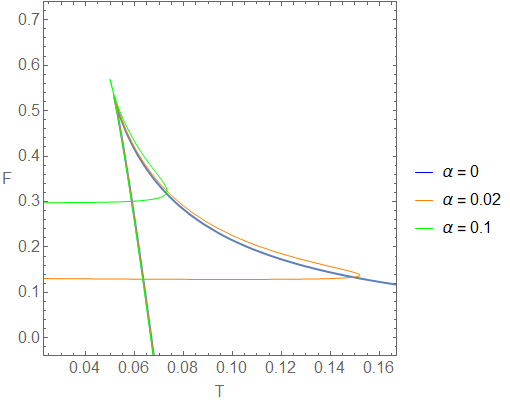}\quad\;\;\;\;\includegraphics[width=0.49\textwidth]{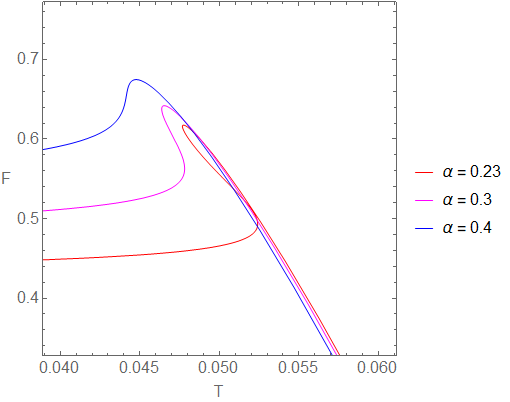}
	\newline
	\includegraphics[width=0.54\textwidth]{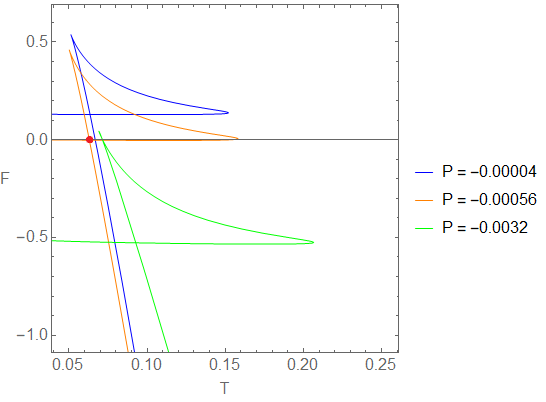}\includegraphics[width=0.54\textwidth]{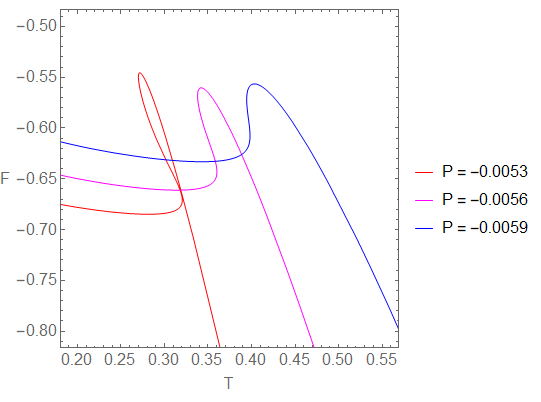}
	\caption { Free energy of the uncharged 4D Gauss-Bonnet-de Sitter black hole  with $r_c=4$. Hawking-Page phase transitions from radiation to a large black hole occur where the free energy crosses $F=0$. \textbf{Top:} Varying Gauss-Bonnet coupling with $P=-0.00004$. The blue curve corresponds to the Einstein limit. On the left, small-large phase transitions occur where the free energy self-intersects. On the right, large Gauss-Bonnet coupling causes the free energy to become single-valued, though this region is above $F=0$ and thus unstable. \textbf{Bottom:} Varying pressure with $\alpha=0.02$. A triple point (red dot) occurs when the pressure is large enough in magnitude. Above this critical pressure we observe a small-large phase transition, as in the green curve on the left. On the right, large negative pressures cause the phase transition to become zeroth-order (pink/blue curves). This region is below $F=0$ and thus stable. 
	}
\label{fig:Fig3}
\end{figure}

Having identified regions in the parameter space where phase transitions may occur, we now turn our attention to the free energy $F$ of the system. In \figurename{ \ref{fig:Fig3}} we plot the free energy $F=E-TS$ parametrically as a function of $T$ with $r_+$ as the parameter. To begin with, we note that, since the free energy contains terms involving $\phi$, we must numerically integrate our solution \eqref{phi} as no closed form of this integral exists. The thermodynamically preferred state is the one that globally minimizes the free energy. As the temperature of the system increases, the system will follow the path with lowest free energy whenever a crossing is reached.
\\

 In the top of \figurename{ \ref{fig:Fig3}}, we see that for non-zero $\alpha$ a ``swallowtail" shape emerges where the free energy of the black hole intersects with itself (as in the green and yellow curves). This signals a first-order small-to-large black hole phase transition. These crossings however occur above $F=0$, so this region is unstable and in fact only a Hawking-Page transition occurs where the lower arm of each curve eventually crosses $F=0$. This behaviour is qualitatively similar to that of 5-dimensional Gauss-Bonnet gravity, which is itself distinct from its higher dimensional generalizations \cite{haroon2020}. As $\alpha$ increases further,  the right-hand terminal point of  the swallowtail moves left.  Eventually, a critical value is reached (the red curve on the top right) where swallowtail pinches off into a ``loop". Beyond this point, there is a zeroth-order small-large phase transition where the free energy jumps discontinuously as the temperature increases (as in the pink curve), accompanied by a non-zero latent heat. When $\alpha$ is larger still (blue curve) the free energy becomes single-valued and the transition disappears. Like the figure on the left, this occurs above $F=0$ so in fact only the Hawking-Page transition exists. 
\\

On the bottom of \figurename{ \ref{fig:Fig3}}, we plot the free energy at fixed value of the Gauss-Bonnet coupling and varying pressure. We see that the effect of increasing the (magnitude of the) pressure is qualitatively similar to that of the coupling constant, with the notable distinction that  larger (more negative) pressures cause a global decrease in the free energy. Thus there is a critical pressure (the orange curve) where the small-large transition becomes stable and dominates over the Hawking-Page transition seen in the upper left of \figurename{ \ref{fig:Fig3}}. As the pressure becomes increasingly negative, the right-hand terminal point of the swallowtail moves left, eventually intersecting the left part of the curve and pinching it into a loop. At still more negative pressures, we again see a zeroth-order phase transition, this time with negative free energy. Again, similar to what was observed for 5D Gauss-Bonnet de Sitter black holes, we see that the region of $(\alpha,P)$-space where a first-order phase transition exists is compact, though not bounded by two second-order critical points. Finally, we note that a triple point occurs at a critical pressure (the orange curve in the bottom left of \figurename{ \ref{fig:Fig3}}), where the small black hole, large black hole, and radiation phases have the same free energy and coexist at the same temperature (marked by a red dot). This novel triple point was recently observed for exotic Lovelock black holes in six dimensions \cite{hull2021}.
\\
\begin{figure}[h]
	\includegraphics[width=0.52\textwidth]{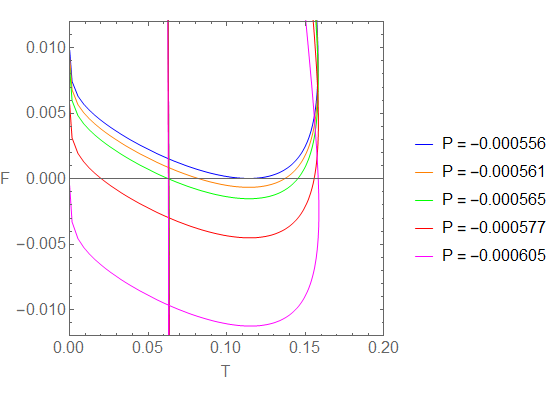}\quad\includegraphics[width=0.49\textwidth]{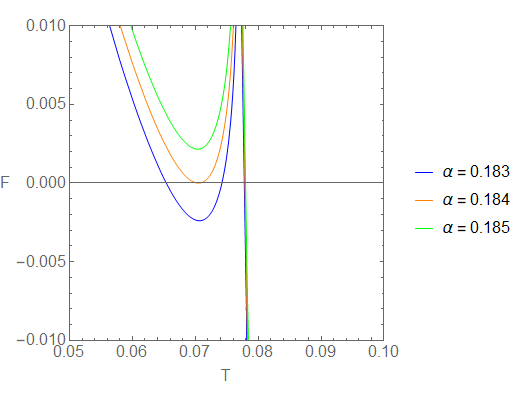}
	\caption{Free energy of the 4D Gauss-Bonnet-de Sitter black hole  with $r_c=4$, with the vertical axis rescaled to show phase structure in a narrow range of parameters about which the curve crosses $F = 0$ \textbf{Left:} Varying pressure with $\alpha = 0.02$. 
	 \textbf{Right:} Varying Gauss-Bonnet coupling with $P= -0.0294$.
	}
\label{fig:Fig4}
\end{figure}

Finally, in \figurename{ \ref{fig:Fig4}} we plot the free energy as a function of equilibrium temperature, rescaled to show the behaviour   in a narrow range of parameters near those where the lower arm of the curve crosses the horizontal axis (near the yellow curve at the bottom left of \figurename{ \ref{fig:Fig3}}. On the left, we show the result of varying the pressure around this intersection for small $\alpha$ and $P$ (i.e. small enough that a swallowtail exists). We see that a range of phase structures are possible depending on small changes to the pressure. For large pressure (as in the pink line), the small black hole phase persists down to $T=0$, and a small-large transition occurs at a certain critical temperature. As the pressure decreases in magnitude, part of the small black hole branch crosses the $F=0$ line and becomes unstable (as in the red line). Here, there are two critical temperatures corresponding to a transition from radiation at low temperatures to a small black hole, followed by a small-large transition at a second critical temperature. When the pressure reaches the critical value of $P\sim 5.65\!\times\!10^{-4}$, the small-large transition point has vanishing free energy, such that the radiation, small, and large black hole phases all coexist. This is the triple point noted in the previously, and occurs in the green curve of \figurename{ \ref{fig:Fig4}}. Above this pressure, only Hawking-Page transitions occur as the small black hole branch lies entirely above the $F=0$ line.
\\

On the right of \figurename{ \ref{fig:Fig4}}, we vary the Gauss-Bonnet coupling at constant pressure, choosing values such that the curve self-intersects (as in the red curves in the right of \figurename{ \ref{fig:Fig3}}). The situation is qualitatively similar to the previous one, except that the large black hole branch (the one that extends down towards $F=-\infty$) now lies entirely to the right of the small black hole branch. Above $\alpha=0.184$, we have only a Hawking-Page transition (green curve) where the lower arm crosses $F=0$. Below $\alpha=0.184$ a new reentrant phase transition appears (blue curve), where increasing the temperature causes the following sequence of phases to appear: radiation-small black hole-radiation-large black hole. This kind of reentrant radiation-black hole-radiation transition has only been seen previously in Born-Infeld-de Sitter black holes in the grand canonical (fixed potential) ensemble \cite{simovic2019}. Further exploration is warranted to uncover how these transitions are related, since it is curious that the the same structure appears in different thermodynamic ensembles when a scalar field is present. 

\section{Charged 4D Gauss-Bonnet Black Holes}\label{charged}

We now consider the phase structure of charged black holes, i.e. the inclusion a $U(1)$ Gauge field in the action \eqref{fullact}. The action in terms of the metric function and scalar field is the same as before for the reasons already discussed. Hence the energy, entropy, free energy, and temperature are all take the same functional form as before, with the caveat that the metric function is now given by
\be
f(r) = 1 + \frac{r^2}{2\alpha}\left(1 - \sqrt{1 + \alpha\left(\frac{4\Lambda}{3} + \frac{8 M}{r^3}- \frac{4 Q^2}{r^4}\right) } \right)\ .
\ee
With this metric function, we can readily repeat the analysis carried out in the previous section and examine the effect of the gauge field on the phase structure. We first remark briefly on how the first law should be appropriately generalized when charge is present.

\subsection{The First Law}

In the charged case, the extended first law differs from \eqref{firstlaw} only by the inclusion of a term associated with a charge variation $dQ$, along with the conjugate potential $\phi$. Unlike the scalar field, which is not a primary charge and not associated with an integration constant, the electric charge $Q$ constitutes primary hair for the black hole and must have its variation accounted for in the first law. We have that:
\be\label{firstlaw2}
dE=TdS+VdP+\sigma dA+\Phi_{GB}\,d\alpha+\psi\,dQ
\ee
Here, $\psi$ is the electric potential as measured by an observer at the surface of the cavity (at $r=r_c$), and is not fixed in the canonical ensemble. It is the charge $Q$ that is fixed instead, which has the consequence of preventing a charged black hole from transitioning into empty space via a Hawking-Page transition\footnote{One can instead consider a canonical ensemble where $\psi$ is fixed instead of $Q$ and explore the consequences, as was done in \cite{simovic2019}. In such an ensemble, a black hole would evaporate into charged radiation, which allows the potential to stay fixed.}. This is due to the fact that a charged black hole must have $M^2>Q^2$, and so a lower bound on the allowed mass appears when fixing $Q$. This means that the unstable regions (above $F=0$) observed in the uncharged case may now dominate the statistical ensemble, since the empty ($M=0$) spacetime is no longer accessible by the black hole system.
\subsection{Phase Structure of Charged Black Holes}

As before, we begin by looking at plots of the temperature as a function of the event horizon radius. First, we note that the effect of changing either the coupling constant or pressure has an effect perfectly analogous to that in the uncharged case, so that increasing either sufficiently will make the temperature a one-to-one function of $r_+$. 
\\
\begin{figure}[h]
\centering\includegraphics[width=0.6\textwidth]{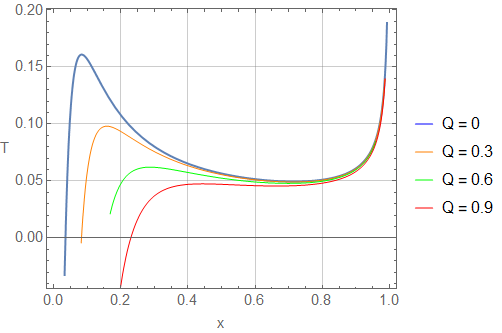}
	\caption{Temperature of the charged 4D Gauss-Bonnet-de Sitter black hole  as a function of $x = r_+/r_c$ and varying $Q$ with $r_c=4$, $\alpha = 0.02$, $P=-0.00080 $. We note that increasing the charge bears a striking resemblance to increasing the coupling constant, with large enough leading to the temperature becoming one-to-one as a function of $r_+$.
	}
	\label{fig:Fig5}
\end{figure}

\figurename{ \ref{fig:Fig5}} shows such a plot of temperature with fixed $\alpha$, $P$ and varying $Q$. We notice that the effect of changing $Q$ is very similar to that of changing $\alpha$. If either of these three parameters is large enough, the temperature will become a monotonic function of $r_+$, excluding the possibility of first order phase transitions. To better understand this in the context of the free energy of these phases, we parametrically plot the free energy as a function of temperature as in the uncharged case.
\\

\begin{figure}[h]
\includegraphics[width=0.49\textwidth]{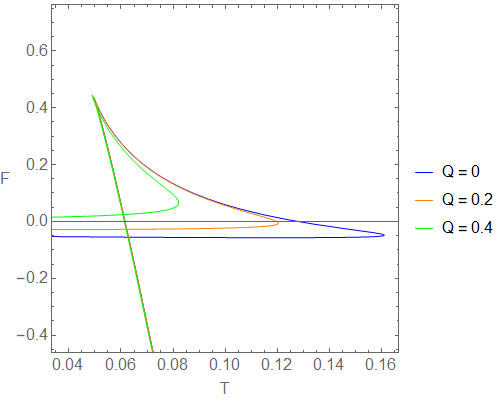}\quad\quad\includegraphics[width=0.49\textwidth]{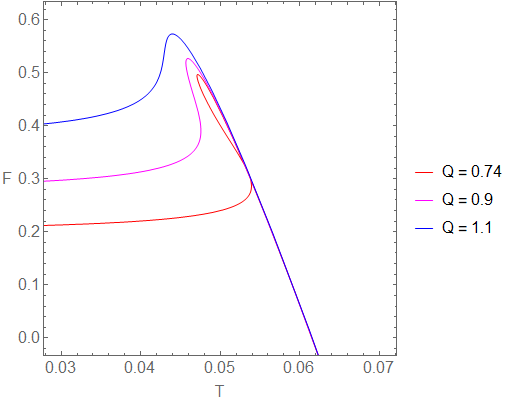}
\newline
	\includegraphics[width=0.53\textwidth]{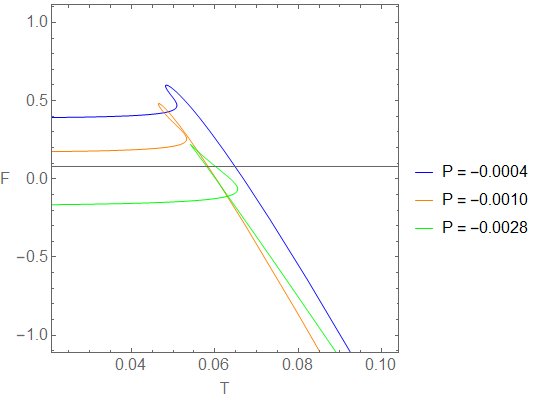}\includegraphics[width=0.55\textwidth]{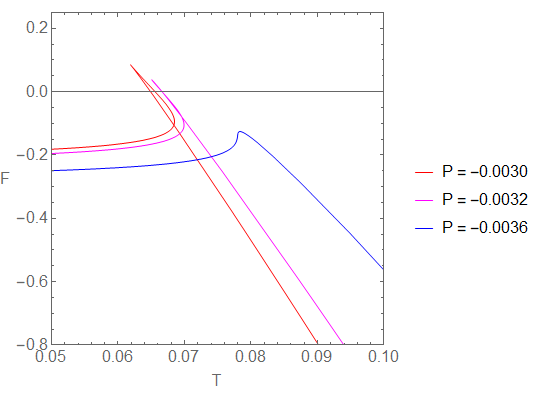}
	\caption{Free energy of the charged 4D Gauss-Bonnet-de Sitter black hole against temperature. \textbf{Top:} Varying $Q$ with $r_c=4$, $\alpha = 0.02$, $P=-0.00080 $. Note that varying $Q$ is qualitatively very similar to varying $\alpha$ in the uncharged case; the curves (and their interpretation) are qualitatively identical to the top of \figurename{ \ref{fig:Fig3}}. \textbf{Bottom:} Varying $P$ with $r_c=4$, $\alpha = 0.01$, $Q=0.8$. We note the existence of a "swallowtube" for this choice of parameters; i.e. there is a minimum and maximum pressure between which small-large phase transitions occur.
	}
\label{fig:Fig6}
\end{figure}
On the top of \figurename{ \ref{fig:Fig6}} we show such a plot, again using $r_+$ as a parameter. We see a close similarity to the plots varying $\alpha$ from the uncharged case, but with a different interpretation. Because we are working in the canonical (fixed charge) ensemble, the radiation phase is inaccessible to the system when $Q>0$, because a black hole cannot evaporate completely while retaining a nonzero charge. For large enough $Q$, the self-intersection of the free energy line disappears, being replaced by a zeroth-order phase transition where the system undergoes a discontinuous jump in $F$ with a nonzero latent heat. At sufficiently large $Q$ the free energy becomes single-valued for all $T$ and no phase transitions occur. Finally, we note that the small-scale behaviour near where the lower arm of the swallowtail crosses $F = 0$ shown in \figurename{ \ref{fig:Fig3}} in the uncharged case is unaltered in the charged case. On that scale too, increasing $Q$ has an effect essentially equivalent to that of increasing $\alpha$. On the bottom we show the effect of varying pressure at fixed charge and (small) $\alpha$. For certain values of $Q$ and $\alpha$, there is a finite range of pressures bounded by a minimum and maximum within which a small-large transition is present. This ``swallowtube" phenomenon is known to exist in the charged Schwarzchild-de Sitter case \cite{simovic2019a}; above a certain $\alpha$ (which depends on each particular choice of $Q$) it vanishes. Distinct from that case is the fact that, whereas the swallowtube terminates at two second-order critical points for de Sitter black holes in Einstein gravity, here the tube terminates at a minimum (yellow curve) and maximum (pink curve) pressure in a zeroth-order phase transition. The swallowtube structure is known to arise generically in de Sitter space due to the presence of the cosmological horizon \cite{simovic2021b}. For the first time, we observe zeroth-order phase transitions bounding this tube, which are ordinarily second-order critical points. It will be important to examine in the future how modifying the structure of the theory at the level of the Lagrangian alters the nature of these critical points, which until now appeared to be robust in that they remained second-order even in higher dimensions and when e.g. non-linear extensions of Maxwell electrodynamics are included.

\section{Conclusions}\label{conclusions}

We have examined for the first time the thermodynamics and phase structure of asymptotically {\it de Sitter} ($\Lambda>0$) black hole solutions in the recently formulated 4D Gauss-Bonnet gravity. Alongside Schwarzschild- and Kerr-de Sitter black holes in Einstein gravity, these solutions (especially their rotating extensions) represent perhaps the most important models of astrophysical black holes, which are almost certainly (at least approximately) asymptotically de Sitter. By investigating their thermodynamic properties, we have not only revealed how phenomena present in AdS black holes and in other theories persist in this new theory of gravity, but also demonstrated the unique effect the presence of a scalar field has in the thermodynamic description. This was seen also in \cite{simovic2021}, where coupling Einstein gravity to a scalar field altered the phase structure of black hole solutions in a matter very different from the inclusion of angular momentum, charge, or higher dimensions.
\\

Remarkably, uncharged black holes in 4D Gauss-Bonnet gravity display a wide variety of phenomena that ordinarily only appear in very high dimensions in AdS spacetimes \cite{frassino2014}. We observe Hawking-Page transitions, small-to-large black hole phase transitions, as well as the occurrence of triple points where radiation coexists with the small and large black hole phases. We also observe a unique zeroth-order phase transition occurring both in the charged and uncharged cases for certain values of the charge and/or Gauss-Bonnet coupling. Finally, we observe the typical ``swallowtube'' behaviour seen in a variety of other asymptotically de Sitter contexts. Such a tube represents a compact (in pressure) region in the parameter space within which small-large phase transitions occur, and is unique to de Sitter owing to the presence of the cosmological horizon. In this case however, instead of the minimum ($P_{min}$) and maximum ($P_{max}$) pressures corresponding to second-order critical points with no transitions occurring outside this range, we instead observe zeroth-order phase transitions outside of the range $[P_{min},P_{max}]$. This surely is a consequence of the Horndeski-type theory we are considering here, and it will be interesting to examine further how exactly these transitions are associated with the presence of the scalar field $\phi$ in this 4D Gauss-bonnet theory.
\\

All said, we have demonstrated that black hole solutions in the recently formulated 4D Gauss-Bonnet theory of gravity have a well-defined thermodynamic interpretation even in the asymptotically de Sitter case, despite difficulties associated with the presence of the cosmological horizon. By adopting a Euclidean path integral approach and fixing boundary-value data on a finite-radius surface we are able to study the equilibrium thermodynamic properties of these black hole solutions. We have shown that many of the thermodynamic features common in AdS black holes (triple points, Hawking-Page transitions, etc.) appear also in 4D Gauss-Bonnet gravity, alongside some features which are known to arise only in de Sitter spacetimes (like the swallowtube). The presence of the scalar field here has the interesting consequence of supporting much more exotic structure (like reentrant phase transitions) than would be otherwise expected in $D=4$, as high dimensionality is generally required to observe such features.

\section{Acknowledgements} This work was supported in part by the Natural Sciences and Engineering Research Council of Canada.

\section*{Appendix A}

Here we present expressions for the conjugate thermodynamic potentials $\{V,\psi,\sigma,\Phi_{GB}\}$ as determined by the first law \eqref{firstlaw2} for the 4D Gauss-Bonnet black hole. The electric potential $\psi$ is defined only in the charged case, which differs from the uncharged case only by the replacement of $f$ and $f_0$ by their appropriate $Q$-dependent versions in the expressions below. $\partial_{r_c}$ indicates a partial derivative with respect to $r$, evaluated at $r=r_c$. We find:

\begin{align}
 V&=8 \pi  \Bigg[\frac{r_c^3}{6}-\frac{r_c^3-r_+^3}{6 \sqrt{f}}-\frac{1}{2} r_c\,\partial_{\Lambda} f_0+\frac{\partial_{\Lambda}f \left(\sqrt{f} \Lambda r_c^3+3 (f-1) r_c-\Lambda r_+^3+3 r_+\right)}{12 f^{3/2}}\nonumber\\
 &+\alpha  \Bigg(\frac{\partial_{r_c} f \left(\phi\,\partial_{\Lambda}f-2 f \partial_{\Lambda}\phi\right)-2 f \phi\,  (\partial_{\Lambda}\partial_{r_c} f)}{2 f^{3/2}}-\frac{\partial_{\Lambda}f \left(\frac{3 f^2 r_+-r_c+r_+}{f^{3/2} r_+}-4\right)}{4 r_c}+\frac{\left(2 f_0^{3/2}-3 f_0+1\right) \partial_{\Lambda}f_0}{2 \sqrt{f_0}\, r_c}\nonumber\\
 &+(\partial_{r_c}f_0)(\partial_{\Lambda}\phi)-\phi_0 (\partial_{r_c}\partial_{\Lambda} f_0)\Bigg)\Bigg]\nonumber \tag{A.1}
\end{align}

\begin{align}
	\sigma&=\dfrac{1}{8\pi r_c}\left[\frac{1}{2}\left(1-r_c^{2} \Lambda-\sqrt{f}+f_0\right)-\frac{1-r_c^{2} \Lambda+\frac{1}{2} r_c \partial_{r_c} f}{2 \sqrt{f}}+\frac{3 \partial_{r_c} f(r_c-r_+)-\left(r_c^{3}-r_+^{3}\right) \Lambda \partial_{r_c} f}{12 f^{3 / 2}} \right.\nonumber\\
	&\left.+\ \frac{1}{2} r_c \partial_{r_c} f_0 +\alpha\left( 
	\frac{\frac{1}{\sqrt{f}}+2 f-f^{3 / 2}+2 \sqrt{f_{0}}-2 f_{0}^{3 / 2}+f_{0}^{2}-3}{2 r_c^{2}}-\frac{\partial_{r_c} f+2 r_+ \phi f^{\prime 2}}{4 r_+ f^{3 / 2}}+\frac{\partial_{r_c} f \phi^{\prime}}{\sqrt{f}}-\partial_{r_c} f_0 \phi_{0}^{\prime}   \right. \right. \nonumber\\
	&\left. \left.  +\ \frac{\frac{\partial_{r_c} f}{f^{3 / 2}}-\partial_{r_c} f+3 \sqrt{f} \partial_{r_c} f-\frac{2 \partial_{r_c} f_0}{\sqrt{f_{0}}}+(6 \sqrt{f_{0}}+f_0) \partial_{r_c} f_0}{4 r_c}  +\frac{\phi\, \partial_{r_c}^2 f}{\sqrt{f}}-\phi_{0}\, \partial_{r_c}^2 f_0 \right)\right]	 \tag{A.2}
\end{align}

\begin{align}
    \Phi_{GB}&=\tfrac{1}{2} r_c \partial_{\alpha} f_{0}+\frac{3-f+f^{3 / 2}-2 \sqrt{f_{0}}+2 f_{0}^{3 / 2}-f_{0}^{2}}{2 r_c}+\frac{3(r_c-r_+) \partial_{\alpha} f-\left(r_c^{3}-r_+^{3}\right) \Lambda \partial_{\alpha} f}{12 f^{3 / 2}}\nonumber\\
    &+ \frac{(r_c-r_+)}{2 r_c r_+ \sqrt{f}}-\frac{r_c \partial_{\alpha} f}{4 \sqrt{f}}-\phi_{0} \partial_{r_c} f_{0}-\alpha\left[\frac{(r_c-r_+) \partial_{\alpha} f}{4 r_c r_+ f^{3 / 2}} +\frac{\phi \partial_{\alpha} f \partial_{r_c} f}{2 f^{3 / 2}}+\partial_{\alpha} \phi_{0} \partial_{r_c} f_{0}+\phi_{0} \partial_{\alpha} \partial_{r_c} f_{0} \right.\nonumber\\
   &\left.-\frac{\partial_{\alpha} \phi \partial_{r_c} f+\phi \partial_{\alpha} \partial_{r_c} f}{\sqrt{f}}-\frac{\frac{3}{4} \sqrt{f} \partial_{\alpha} f-\partial_{\alpha} f-\frac{\partial_{\alpha} f_{0}}{2 \sqrt{f_{0}}}+\frac{3}{2} \sqrt{f_{0}} \partial_{\alpha} f_{0}-f_{0} \partial_{\alpha} f_{0}}{r_c}\right]\tag{A.3}
\end{align}

\begin{align}
	\psi&=\tfrac{1}{2} r_c \partial_{Q} f_0+\frac{\left(3(r_c-r_+)-\Lambda\left(r_c^{3}-r_+^{3}\right)-3 r_c f\right) \partial_{Q} f}{12 f^{3 / 2}}+\alpha\left[\frac{(3 \sqrt{f}-4) \partial_{Q} f+(6-4 \sqrt{f_0}) \sqrt{f_0} \partial_{Q} f_0}{4 r_c}\right.\nonumber\\
	&\left.-\frac{(r_c-r_+) \partial_{Q} f}{4 r_c r_+ f^{3 / 2}}-\frac{\partial_{Q} f_0}{2 r_c \sqrt{f_0}}-\frac{\phi \partial_{Q} f \partial_{r_c} f}{2 f^{3 / 2}}+\frac{\phi\, (\partial_{r_c} \partial_{Q} f)+(\partial_{Q} \phi) (\partial_{r_c} f)}{\sqrt{f}}-\partial_{Q} \phi_0 \partial_{r_c} f_0-\phi_0 \partial_{r_c} \partial_{Q} f_0\right]\tag{A.4}
\end{align}

\bibliographystyle{ieeetr}
\bibliography{LBIB}

\end{document}